# Numerical study of anomalous diffusion of light in semi-crystalline polymer structures

E G Kostadinova [1], J L Padgett [2, 1], C D Liaw [3, 1], L S Matthews [1], & T W Hyde [1]

[1] CASPER and Department of Physics, Baylor University, Waco, TX 76706, USA

[2] Department of Mathematical Sciences, University of Arkansas, Fayetteville, Arkansas 72701, USA

[3] Department of Mathematical Sciences, University of Delaware, Newark, DE 19716, USA

E-mail: Eva_Kostadinova@baylor.edu, padgett@uark.edu, liaw@udel.edu, Lorin_Matthews@baylor.edu, Truell_Hyde@baylor.edu

**Abstract.** From the spread of pollutants in the atmosphere to the transmission of nutrients across cell membranes, anomalous diffusion processes are ubiquitous in natural systems. The ability to understand and control the mechanisms guiding such processes across various scales has important application to research in materials science, finance, medicine, and energetics. Here we present a numerical study of anomalous diffusion of light through a semi-crystalline polymer structure where transport is guided by random disorder and nonlocal interactions. The numerical technique examines diffusion properties in one-dimensional (1D) space via the spectrum of an Anderson-type Hamiltonian with a discrete fractional Laplacian operator $(-\Delta)^s$, $s \in (0,2)$ and a random distribution of disorder. The results show enhanced transport for $s < 1$ (super-diffusion) and enhanced localization for $s > 1$ (sub-diffusion) for most examined cases. An important finding of the present study is that transport can be enhanced at key spatial scales in the sub-diffusive case, where all states are normally expected to be localized for a (1D) disordered system.

**Keywords:** anomalous diffusion, fractional Laplacian, spectral approach, semi-crystalline polymers

## I. INTRODUCTION

The ubiquity of anomalous diffusion processes observed in laboratory experiments and in nature is well captured by the words of Klafter and Sokolov [1]: "*although these phenomena are called anomalous, they are abundant in everyday life: anomalous is normal.*" Super-diffusion has been observed in the search patterns of animals [2]–[4], the spread of pollutants in the ocean and the atmosphere [5], [6], and particle transport in turbulent plasmas [7]–[9]. Sub-diffusion has been found characteristic of electron transport in amorphous materials [10], gas transport in porous media [11], signal transmission across cell membranes [12], [13], and protein fluctuation patterns [14]. The variety of natural systems exhibiting anomalous diffusion requires the development of generalized mathematical techniques together with appropriate physical interpretations, adapting such techniques to specific problems. In this paper we study, numerically, anomalous diffusion using the Fractional Laplacian Spectral (FLS) method [15], which is suitable for modeling transport guided by nonlocal interactions and random disorder. Although this technique is general (applicable to any Anderson-type Hamiltonian of the form defined in [16]), here we focus on a Hamiltonian appropriate for the problem of light transport in semi-crystalline polymers, where both disorder and nonlocal interactions can occur [17], [18].

The study of light transport in random media and photonic crystals has resulted in the advancement of a large class of groundbreaking imaging techniques, including coherent backscattering [19], [20], dynamic light scattering [21], [22], and optical coherence tomography [23]. Due to the variety and complexity of disordered media examined with these techniques, both classical and quantum formulations have been used to account for the possibility of various processes affecting light transport, such as Anderson localization, non-linear effects, and nonlocal interactions [24]–[27]. Quantum field theoretical methods have been proposed for the characterization of disordered complex media with short laser pulses in an optical coherence tomography setup [28], [29]. These models investigate light transport in the presence of highly non-linear and discontinuous processes, including interference effects and transition to Anderson localization. An interesting question arises in the case of photon transmission through disordered gain or loss media, where the interplay between strong Anderson localization of light and non-linear effects can yield reduction of the gain or loss induced correlation length, even though no truly localized modes are predicted in this regime [30], [31]. An important application of such theoretical methods is the development of random lasers, where multiple scattering of photons from active resonators within the disordered gain media yields amplification without the need to use external feedback or a mirror system [28]. Polymers have been proposed as candidate gain media for random lasers due to the complexity of their molecular structure and inherent network-mediated interactions [32]. Generally speaking, the physical origin of nonlocal interactions in polymers can be attributed to their microstructure and the way branch points or folding processes lead to deviations from a simple 1D chain structure [33], [34] (see Fig. 4 and related discussion in Sec. IIIB).

Motivated by such applications, the present work aims to extend the theory of Anderson localization of light to include many-body nonlocal interactions and to investigate the resulting anomalous diffusion transport in a semi-crystalline polymer medium. Anderson localization describes the absence of diffusion due to lattice impurities (disorder) [35] and has been examined in various systems, including quantum walkers, Bose-Einstein condensates, photon lattices, and granular chains [36]–[40]. A long-standing question in the theory of Anderson localization is the effect of many-body nonlocal interactions on diffusion in disordered media [41]. Many-body transport has been investigated in fermionic systems at high temperatures, quantum spin chains, and weakly-correlated electrons [42]–[44]. The interplay between stochastic disorder and nonlocal interactions can lead to the excitation of various dynamical instabilities, such as turbulence in plasma [8], as well as solitary waves and solitons in soft matter [45], [46]. Thus, generalizing the theory of Anderson localization to account for nonlocality may address important questions beyond the realm of condensed matter physics.

In this paper, we use results from spectral theory and fractional calculus to investigate, numerically, anomalous diffusion of light due to nonlocal interactions in a semi-crystalline polymer medium with Anderson-type random disorder. One of the main novelties of the spectral approach is that it represents a different way of approaching the localization problem: the spectral method does not assume the (commonly accepted but not rigorously physical) scaling hypothesis, as discussed in [47], [48]. In particular, long-range interactions do NOT arise from an upscaling effect or procedure, but rather directly from the explicit numerical model as described in [15]. In this approach, nonlocal interactions are modeled using the fractional Laplace operator $(-\Delta)^s, s \in (0,2)$, which is often employed in the study of anomalous diffusion [49], [50]. In the regime $s \in (0,1)$, the fractional Laplacian has been used to demonstrate the emergence of anomalous diffusion and Lévy flights in the dynamics of both networks and lattices with nonlocal interactions [51]–[53]. The related fractional Laplacian matrix of networks was further used to define a fractional random walk and a fractional Schrödinger equation, establishing a fractional formulation of quantum mechanics that combines long-range dynamics with the quantum superposition of states [54]–[56]. Thus, this operator is suitable for both classical and quantum formulations of the diffusion problem.

The properties of the *discrete* fractional Laplacian $(-\Delta)^s, s \in (0,2)$ were examined in detail in our previous work [15], where we also discussed its application in a spectral approach [47] to the Anderson localization problem. The resulting Fractional Laplacian Spectral (FLS) model was employed in one-dimensional numerical simulations [15] which demonstrated qualitative transport enhancement in the super-diffusive case, $s \in (0,1)$, and enhanced localization in the sub-diffusive regime, $s \in (1,2)$. Here we extend the application of the FLS technique by examining transport as a function of disorder concentration, exponent of the Laplacian, and vector scales in the Hilbert space. These parameters are interpreted as physical properties of a polymer medium, such as molecular chain structure and orientation.

In Sec. II, we summarize the main theoretical results used as a basis for the FLS model. Section III introduces the problem of light transport through a semi-crystalline polymer and discusses the physical meaning of the key quantities in the FLS simulation in this context. Specifically, we discuss the choice of 1D geometry, the interpretation of disorder concentration, and the physical origins of nonlocal interactions within polymer media. Section IV presents a numerical investigation of anomalous diffusion for various disorder concentrations, properties of the nonlocal operator, and vector scales in the Hilbert space. A major finding of the present study is the existence of enhanced transport at key vector scales in the sub-diffusive regime, where all states are typically expected to be localized in a 1D disordered system. The possible physical origins of these results are discussed in Sec. V. Sec. VI provides a summary of the results and outlines directions for future research. The details of the numerical simulation are summarized in Appendix A. (Detailed description of the simulation and related proofs can be found in [15].) In Appendix B, we examine numerical instabilities due to roundoff errors and other approximations.

## II. FRACTIONAL LAPLACIAN SPECTRAL MODEL

### A. Random Fractional Discrete Schrödinger Operator

In this paper, we extend the 1D Anderson localization problem to include nonlocal interactions by considering the *random fractional discrete Schrödinger operator*

$$H_{s,\epsilon} := (-\Delta)^s + \sum_{i \in \mathbb{Z}} \epsilon_i \langle \cdot, \delta_i \rangle \delta_i, \tag{1}$$

where $(-\Delta)^s$, $s \in (0,2)$ is the discrete fractional Laplacian, $\delta_i$ is the $i$th standard basis vector of the 1D integer space $\mathbb{Z}$, $\langle \cdot, \cdot \rangle$ is the $\ell^2(\mathbb{Z})$ inner product, and $\epsilon_i$ are independent variables, identically distributed according to a uniform (flat) distribution on the interval $[-c/2, c/2]$, with $c > 0$. The operator in (1) is appropriate for studying particle transport characterized by nonlocal interactions, such as positive or negative correlations in a many-body disordered system. It is expected that in the *super-diffusive* regime $s \in (0,1)$, transport is enhanced due to positive correlations, while in the *sub-diffusive* regime $s \in (1,2)$, propagation through the medium is impeded due to negative correlations [49], [50]. In the limit where $s \to 1$, the fractional Laplacian reduces to the classical case. Note that, in our model, the effect of random disorder is considered as a localization mechanism, which is distinct from the effect of negative correlations in the sub-diffusive regime.

Throughout the literature, there are numerous definitions of the fractional Laplacian involving singular integrals, semigroups of operators, and harmonic extensions [58]. Caffarelli and Silvestre [59] demonstrated that the operator $(-\Delta)^s, s \in (0,1)$ can be constructed from a harmonic extension problem to the upper half space via an operator that maps Dirichlet to Neumann boundary conditions. Recently, Chen et al. [60] derived similar extension results for the case $(-\Delta)^s, s > 1$. Starting with a hyper-singular integral form of the fractional Laplacian, Padgett *et al.* [15] combined these techniques to obtain the following 1D series representation of $(-\Delta)^s$ for the interval $s \in (0,2)$

$$(-\Delta)^s u(n) = \sum_{m \in \mathbb{Z}; m \neq n} \bigl(u(n) - u(m)\bigr) K_s(n-m), \qquad (2)$$

where

$$K_s(m) = \begin{cases} \dfrac{4^s \Gamma(1/2 + s)}{\sqrt{\pi} |\Gamma(-s)|} \cdot \dfrac{\Gamma(|m| - s)}{\Gamma(|m| + 1 + s)}, & m \in \mathbb{Z} \setminus \{0\}, \\ 0, & m = 0. \end{cases} \qquad (3)$$

Here $u$ is a discrete function on $\mathbb{Z}$ and $\Gamma$ is the Gamma function. Numerical simulations by Padgett *et al.* [15] provided preliminary confirmation that the representation in (2) yields enhanced transport (super-diffusion) for $s \in (0,1)$ and enhanced localization (sub-diffusion) for $s \in (1,2)$ when compared to the classical case $s = 1$. The present work expands on this study by investigating how transport at different 1D spatial scales depends on the interplay between disorder concentration and the characteristics of the nonlocal interactions.

### B. Physical Interpretation of $s$ and $c$

A detailed discussion on the physical interpretation of the exponent $s$ can be found in Sec. 2.2. of [15] and in [61]. Here we summarize the basic logic of this interpretation. In the classical case, where $s = 1$, the function $u$ from equation (2) is a discrete harmonic function on the space $\mathbb{Z}$, that is $(-\Delta)u = 0$. The discrete harmonic function $u$ represents a physical situation in which a particle is equally likely to "jump" to either one of its nearest neighbors under the action of the discrete Laplacian. In other words, in 1D space $u$ satisfies the following mean value property

$$u(n) = P(n+1)u(n+1) + P(n-1)u(n-1) = \frac{1}{2}u(n+1) + \frac{1}{2}u(n-1), \qquad (4)$$

where $P(n+1)$ and $P(n-1)$ are the probabilities that a particle located at the lattice point $n$ will jump to the lattice points $n+1$ or $n-1$, respectively. Following the same logic, assume that the function $u$ is a fractional discrete harmonic function, that is, $(-\Delta)^s u = 0$. Then from equation (2), $u$ satisfies the following fractional mean value property

$$u(n) = \sum_{m \in \mathbb{Z}; m \neq n} u(m) P_s(n-m) = \sum_{m \in \mathbb{Z}; m \neq n} \frac{u(m) K_s(n-m)}{K_s(m)}, \qquad (5)$$

where $P_s(n-m)$ is the probability that a particle located at the lattice point $n$ will jump to *any* other lattice point $m$ in the space $\mathbb{Z}$. From the definition of $K_s(m)$ in equation (3), one can show that a fraction $s \in (0,1)$ yields increased probability for particle transport away from its initial position when compared to the classical case $s = 1$. In the $s \in (1,2)$ case, while the time-evolved dynamics is likely to enhance particle localization, as compared to the $s = 1$ case, the particle transport in this regime is still guided by the action of the nonlocal operator $(-\Delta)^s, s \in (1,2)$. The resulting process can be thought of as a superposition between the action of the classical Laplacian plus the action of a fractional Laplacian with exponent $\tilde{s} \in (0,1)$, i.e., the sub-diffusive operator can be decomposed into $(-\Delta)^s = (-\Delta)^{\tilde{s}}(-\Delta)$. When the two actions are combined together, there is an increased probability that the particle jumps back to the original position. This point will be further discussed in Sec. V, where we interpret the results from the numerical simulation.

Generally speaking, the operator in (1) describes a lattice with atoms located at the integer lattice points in $\mathbb{Z}$. The random disorder part in (1), given by the variables $\epsilon_i \in [-c/2, c/2]$, can be interpreted as fluctuations of the lattice on-site potential energies, i.e., energy varies randomly from site to site. Such fluctuations can be physically caused by different mechanisms, including random spacing of impurities, random displacements of atoms from the geometrical lattice points, or random arrangements of electronic or nuclear spins. In the present study of light transport in a semi-crystalline polymer, the randomness describes fluctuations in the molecular orientation due to folding of the polymer chains (further discussed in Sec. III). As the disorder concentration $c$ is increased beyond a certain threshold value, diffusion comes to a halt, which is commonly referred to as Anderson localization [35]. For the one-dimensional case, in the nearest-neighbor (local) approximation, it is known that Anderson localization is induced by any nonzero concentration of disorder [57]. However, in the nonlocal approximation studied here, the localization behavior can be either enhanced or suppressed by varying the fractional exponent $s$ on the Laplacian in equation (1). This allows the identification of qualitative differences between classical and anomalous diffusive transport and will be used as a criterion for the interpretation of numerical results (see Appendix A for details).

## C. Spectral Approach to Transport

In our model, the time evolution of the initial state $\varphi_0$ of the system is generated under the iterative application of the Hamiltonian $H_{s,\epsilon}$ in (1). For convenience, let $\varphi_0 \equiv \delta_0$, where $\delta_0$ is located at the origin of the 1D integer space $\mathbb{Z}$. Features of the resulting transport behavior are then determined from a numerical test, called the spectral approach, where one computes the mathematical distance between the time evolution sequence of the initial state $\delta_0$ and any other fixed state $\nu$ in the same Hilbert space, such that $\nu \neq \delta_0$. The time evolution of $\delta_0$ under the action of $H_{s,\epsilon}$ is given by the sequence $\{\delta_0, H_{s,\epsilon}\delta_0, H_{s,\epsilon}^2\delta_0, \ldots, H_{s,\epsilon}^\tau\delta_0\}$, where $\tau \in \mathbb{N}$ is the number of timesteps (equivalently, $\tau$ is the number of iterations of the Hamiltonian). Let $\{\varphi_0', \varphi_1', \varphi_2', \ldots, \varphi_\tau'\}$ be the sequence of $\ell^2(\mathbb{Z})$ orthogonal vector states obtained from Gram-Schmidt orthogonalization of the original sequence $\{\delta_0, H_{s,\epsilon}\delta_0, H_{s,\epsilon}^2\delta_0, \ldots, H_{s,\epsilon}^\tau\delta_0\}$. This step allows for a proper definition of a mathematical distance in the Hilbert space. Then, for any nontrivial vector $\nu \neq \delta_0$, we define the distance parameter (mathematical distance) as

$$D_{s,\epsilon}^\tau := \sqrt{1 - \sum_{k=0}^{\tau}\left(\frac{\langle \nu, \varphi_k'\rangle}{\|\nu\|\|\varphi_k'\|}\right)^2}, \tag{6}$$

where $\varphi_k'$ is the $k$th term of the sequence $\{\varphi_0', \varphi_1', \varphi_2', \ldots, \varphi_\tau'\}$. Here, $\langle \cdot, \cdot \rangle$ is the $\ell^2(\mathbb{Z})$ inner product and $\|\cdot\|^2 = \langle \cdot, \cdot \rangle$. Equation (6) was originally derived in Liaw [47], where results from spectral theory were used to verify the following conjecture:

**Extended States Conjecture:** *For an Anderson-type Hamiltonian, if one can find a nontrivial vector $\nu$, for which the limit of the distance parameter approaches a positive value as time* *approaches infinity, then the spectrum of the Hamiltonian includes an absolutely continuous (ac) part [1], which indicates the existence of extended energy states.*

In other words if one demonstrates with positive probability that

$$\lim_{\tau \to \infty} D_{s,\epsilon}^{\tau} > 0 \,, \tag{7}$$

then the time-evolved transport behavior of the system under the action of the examined Hamiltonian includes extended energy states. This spectral analysis was previously used to demonstrate numerically the existence of extended states in two-dimensional (2D) lattices of various geometries at small disorder in the nearest-neighbor (local) approximation [63]–[66]. Here we use the spectral method to investigate how the transport behavior in 1D media with random disorder is affected by nonlocal interactions. Since it is a well-known result that for $s = 1$ in 1D all energy states are localized in the presence of disorder [57], examination of the same problem using both $s = 1$ and $s \neq 1$ allows us to identify how nonlocality affects the energy spectrum. This choice of 1D geometry is also appropriate for the problem of light transport through a semi-crystalline polymer due to the discrete nature of photon interactions, the 1D nature of molecular chains in polymers, and the presence of network-mediated interactions originating from chain folding. The extension of the Fractional Laplacian Spectral model to transport problems in 2D media with random disorder and nonlocal interactions is a topic of our future work.

### III. LIGHT TRANSPORT THROUGH A SEMI-CRYSTALLINE POLYMER

As semi-crystalline polymers are characterized both by disorder and nonlocal interactions [17], [18], the operator in (1) is appropriate for describing the Hamiltonian structure of such materials. Light transport in disordered media is characterized by a multiple-scattering process resulting from random fluctuations of the refractive index in space [68]. The addition of nonlocal interactions can yield anomalous diffusion of light, which has been experimentally observed in heterogeneous dielectric materials [69] and in hot atomic vapors [70]. Although polymers are often highly disordered, network-mediated nonlocal interactions can naturally arise from the intricate folding of their molecular chains (e.g., protein folding [71]) or can be artificially induced by doping (e.g., organic semiconductors [72]). Figure 1a shows a schematic representation of a semi-crystalline polymer structure, which exhibits both disorder and crystalline regions of characteristic scale.

A fundamental topic in the study of both artificial polymers and biopolymers is understanding how the molecular structure and orientations of polymer chains determine the macroscopic properties of the material. The molecular orientation within a polymer sample can be determined by measuring its birefringence properties. Birefringence is a phenomenon where an incident light beam is split into two mutually orthogonal rays, called ordinary ray and extraordinary ray (Fig. 1b), which travel with different velocities inside the birefringent material. The ordinary ray travels through the material with the same velocity in every direction, as it would have traveled in an isotropic crystal, while the extraordinary ray travels with a velocity dependent on the propagation direction within the crystal. Birefringent spectroscopy (Sec. 2.5.4 of [67]) is an optical technique in which the sum of the polarizability of all molecular chains within the polymer yields a measurable retardation between the ordinary and extraordinary rays produced by light transmission through a sample material. In this section, we provide a physical interpretation of equation (6) in the context of such retardation between orthogonal components.

---

[1] The spectrum of a Hamiltonian $H$ consists of: (i) an absolutely continuous part, corresponding to extended states and (ii) a singular part, which includes discrete eigenvalues and poorly behaved transitional states (called singular-continuous part of the spectrum). If the spectrum of $H$ coincides with the singular part (i), transport in the examined problem is localized. In the presence of non-vanishing absolutely continuous part of the spectrum, de-localization occurs in the form of extended states (by the RAGE theorem, see e.g., Sec. 1.2 of [62]).

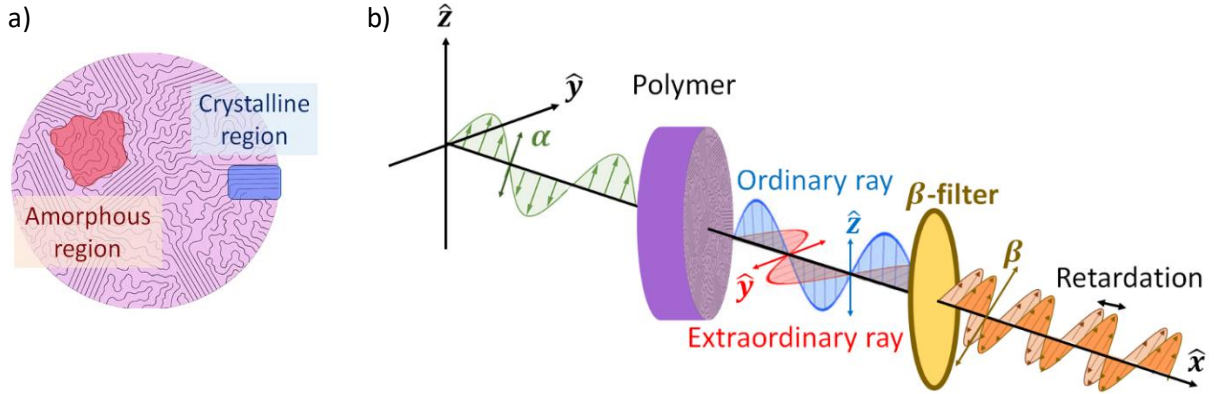

Figure 1. a) Schematic representation of semi-crystalline polymer showing regions of ordered and disordered structures, consisting of folded molecular chains. b) Birefringent spectroscopy: Light, initially polarized at an angle $\alpha_0$ is transmitted through a polymer sample with birefringent properties. The resulting ordinary and extraordinary rays are then passed through an analyzer filter with a fixed

## A. Formulation of the Problem

Here we consider a problem where a plane-polarized light beam is transmitted through a semi-crystalline polymer, and then passed through another polarization filter (Fig. 1b). Assume that the initial polarization direction $\alpha$ of the beam changes due to successive interactions of the photons with the spatial structures within the lattice. A photon-based description of this process requires discretization and probabilistic interpretation, such as the one provided by Dirac in Sec. 2 of [73]. Dirac proposed that light plane-polarized in a certain direction, say at an angle $\alpha$ with respect to the axis $\hat{z}$, can be described as consisting of photons, each polarized in the same direction $\alpha$. When $\alpha$-polarized light passes through a single slit, aligned with $\hat{z}$, the probability of a photon being transmitted approaches $sin^2\alpha$, while the probability of a photon being blocked approaches $cos^2\alpha$. As the number of photons becomes large, the conservation of probability is ensured by the relation $1 = sin^2\alpha + cos^2\alpha$. These probabilities coincide with the measurable fractions of transmitted and blocked light, which justifies the discretized representation. This problem includes a single contact event for each photon with a slit representing single fixed direction in space.

We generalize Dirac's representation to another discretized problem where the photons interact with many contact points within a polymer medium, and are then transmitted through an analyzer filter structure, which can only transmit light polarized along a fixed subset of directions in space. The question of interest is how photon interactions with the complex molecular structures within the sample affect the probability of light retardation, with respect to the analyzer filter. In the following, the iterative application of the operator in (1) guides the time evolution of light propagation due to successive interactions within contact points in the polymer sample [2]. In this representation, we show that the distance parameter $D_{s,\epsilon}^{\tau}$ in equation (6) now quantifies the probability that a photon is transmitted away from the optical axis, defined by the choice of an analyzer filter.

Specifically, consider the 1D Hilbert space $\mathcal{H}$ whose basis vectors $\delta_i$ represent the lattice points in space along the direction of light propagation where the light beam interacts with the sample molecules. The Hamiltonian structure of the polymer sample is modeled by the operator $H_{s,\epsilon}$ in equation (1), where the random disorder term now corresponds to stochastic fluctuations $\{\epsilon_i\}$ in the polarizability of molecular bonds (e.g., due to spatial defects), while the fractional Laplacian accounts for nonlocal scattering events (e.g., due to folding of the molecular chains, as shown in Fig. 1a). Following the discretized description by Dirac, let the incident light beam consist of photons, each initially polarized in the same direction, and let this light beam enter the sample at an initial contact point in space $\delta_0$. In other words, at the 0th timestep, the initial spatial state is given by the delta vector $\delta_0$ of the 1D lattice, with probability 1. The iterative application of the Hamiltonian $H_{s,\epsilon}$ to the initial state $\delta_0$ yields the time evolution of light transport through the sample, given by the sequence $\{\delta_0, H_{s,\epsilon}\delta_0, H_{s,\epsilon}^2\delta_0, \dots, H_{s,\epsilon}^\tau\delta_0\}$, where $\tau$ is the number of timesteps. Using the Gram-Schmidt procedure, one can orthogonalize the members of this sequence to obtain the sequence $\{\varphi_0', \varphi_1', \varphi_2', \dots, \varphi_\tau'\}$, which now represents the light beam propagation to new contact points $\delta_i$ after each scattering event.

Assume that the semi-crystalline polymer sample contains regions of ordered structures, which behave like uniaxial crystals of characteristic scale. To model such regions, consider a unit vector of the form

$$\hat{v} = \frac{v}{\|v\|} = \frac{1}{L}\sum_{j=1}^{L} \delta_j, \qquad (8)$$

which is a linear combination of $L$ number of basis vectors in the Hilbert space, with equal weights. This vector corresponds to a perfectly ordered 1D lattice with $L$ distinct contact points, which represent a subspace of $L$ independent elementary directions defined by the corresponding base vectors in the Hilbert space. This vector can be viewed as a test lattice (or analyzer filter) probing for the existence of crystalline regions of similar scale within the semi-crystalline polymer (Fig. 2a,b,c). After each contact event (iteration of $H_{s,\epsilon}$), the polarization state changes in such a way that allows photons to travel either along the directions defined by $\hat{v}$ or along the direction orthogonal to $\hat{v}$ with a certain probability. Substituting expression (8) into equation (6) gives

$$D_{s,\epsilon}^{\tau} := \sqrt{1 - \sum_{k=0}^{\tau}\left(\frac{\langle v, \varphi_k'\rangle}{\|v\|\|\varphi_k'\|}\right)^2} = \sqrt{1 - \sum_{k=0}^{\tau} cos^2\alpha_k}, \qquad (9)$$

where $\alpha_k$ is the generalized angle between the vector directions in $\hat{v}$ and the directions perpendicular to $\hat{v}$ at the $k$th timestep of the iteration process. Due to the Gram-Schmidt orthogonalization used to obtain the sequence $\{\varphi_0', \varphi_1', \varphi_2', \dots, \varphi_\tau'\}$ and the normalization in equations (8) and (9), at each timestep, the component $cos^2\alpha_k$ and the component $D_{s,\epsilon}^{k}$ are orthogonal. Thus, in this representation, the distance parameter has analogous role as $sin^2\alpha$ in the Dirac example, which allows for a similar interpretation. For a large number of scattering events (large $\tau$), $\sum_{k=0}^{\tau} cos^2\alpha_k$ represents the probability that the photon belongs to an ordinary light ray that propagates in a perfectly ordered crystalline region of fixed finite size $L$, while $\left(D_{S,\epsilon}^\tau\right)^2 = 1 - \sum_{k=0}^{\tau} cos^2\alpha_k$ quantifies the probability that new directional information is generated away from the fixed region due to the dissimilarity between the idealized reference lattice and the examined polymer structure. The conservation of probability in this case is ensured by the Pythagorean Theorem in the Hilbert space.

---

[2] The use of unitary and Hermitian operators that evolve the polarization state of photons in time is common in the study of birefringent crystals [74].

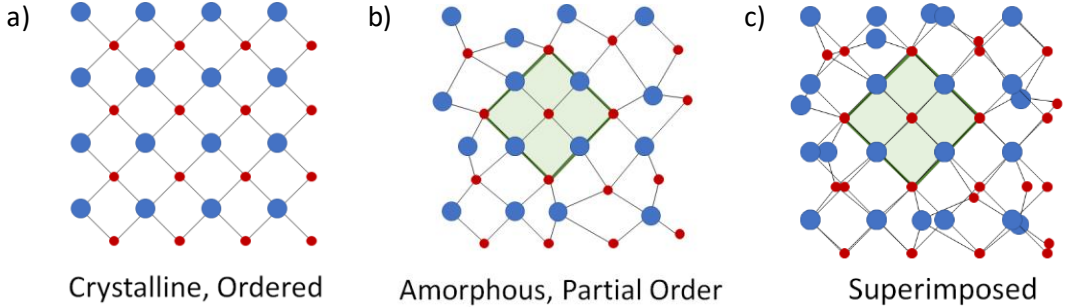

Figure 2. a) Crystalline and b) partially ordered amorphous structures can be c) superimposed to determine the scale of crystalline regions. The distance parameter in equation (6) can be used to quantify the scale of crystalline regions within the semi-crystalline sample.

The plot $D_{S,\epsilon}^\tau(\tau)$ now provides a measure of the difference between spatial directions accumulated under the action of the Hamiltonian $H_{s,\epsilon}$ and the spatial directions in $\hat{v}$, defined by the perfectly-ordered crystalline region of characteristic scale $L$. Therefore, by varying the number $L$ of basis vectors $\delta_j$ in the reference lattice and/or the weights for each $\delta_j$ (fluctuations in the atomic order), one can examine the presence of simpler spatial structures within the more complex polymer medium. If the successive iterations continue to generate new directions as $\tau \to \infty$, the distance parameter $D_{S,\epsilon}^\tau$ is nonvanishing with respect to the reference lattice, indicating dissimilarity between the assumed scale of crystalline regions and the actual structure of the sample.

As discussed in Sec. II, $\lim_{\tau\to\infty} D_{S,\epsilon}^\tau > 0$ indicates the existence of extended states, or transport. Thus, in the present representation, a nonvanishing limiting value of the distance parameter suggests transport beyond the chosen reference lattice scale. In contrast, if $\lim_{\tau\to\infty} D_{S,\epsilon}^\tau = 0$ for a certain choice of the reference lattice $\hat{v}$, the structure of the sample is expected to exhibit regions of crystallinity characterized by size/order, comparable to $\hat{v}$. Based on the above formulation, in Sec. IV we present a numerical study of light transport through semi-crystalline polymer structure described by the Hamiltonian in equation (1). In the present work, we fix the weights on the reference lattice vector from equation (8) and only vary its size for different choices of disorder concentration and properties of the fractional Laplacian. The application of equation (6) when the calculation is performed using a reference lattice with randomly selected vector weights will be explored in our future work.

B. Range of Nonlocality in a Finite Simulation

While the classical Laplacian $\Delta$ models nearest-neighbor (local) interactions, the application of a fractional Laplacian $(-\Delta)^s$ results in interactions that, in principle, extend to infinity. In other words, the analytical expression in equation (2) is exact when the summation is performed over an infinite number of nearest neighbors at each timestep. Due to the finite nature of the numerical simulation, calculation is restricted to a finite number of nearest neighbors, called the effective $range$ of nonlocal interactions, which sets the upper limit for the summation in equation (2). For all cases considered in this work, this limit was varied in the interval $range \in [50, 1000]$ and the corresponding remainder removed by the truncation was $R \sim 1/(range)^2 \sim (10^{-4} - 10^{-6})$. As in all cases the remainder is small, the numerical cutoff yields a good approximation to the fractional Laplacian. See Appendix B for further discussion on the choice of truncation and its possible effect on numerical results.

Introducing a numerical cutoff is not entirely unphysical since nonlocal forces in nature can also exhibit characteristic scales beyond which resulting interactions are not appreciable (for example, consider shielding lengths in plasmas, beyond which electric fields are screened). However, the proper interpretation of numerical results depends on the choice of reference vector size $L$. When $L \lesssim range$, the resulting calculation models a transport problem, where after the first timestep the nonlocal interactions determined from the fractional Laplacian $(-\Delta)^s$ yield a nonzero probability for transport beyond the reference scale examined. Since this probability rapidly increases with the number of timesteps, in these cases the nonlocal interactions are effectively infinite with respect to the fixed scale. In contrast, when $L > range$, a few iterations of the Hamiltonian (timesteps) are needed before the calculation yields a nonzero probability for transport beyond the reference scale examined. Physically, this can be interpreted as shielding of the nonlocal interactions, with the shielding length related to the mismatch between the scales defined by $L$ and $range$.

Initially, we examine cases where the reference vector size is equal to the assumed range of nonlocal interactions, i.e., $L = range$. This intuitive choice makes sense when one considers the folded chains forming the semi-crystalline polymer structure in Fig. 1a and the simplified picture in Fig. 3. While classical diffusive transport occurs only through nearest-neighbor interactions along a single molecular chain, the presence of folding yields the possibility of nonlocal transport across topological neighbors. In case of symmetric folding, characteristic of crystalline regions within the polymer, the nonlocal effects yield enhanced transport (or super-diffusion) for smaller molecular weight (number of molecules per chain). However, larger molecular weight results in increased thickness of the amorphous regions [75]–[78], where the complexity of folds can result in transport retardation (or sub-diffusion). Thus, it is reasonable to expect that a correspondence exists between the characteristic size of crystalline or amorphous regions and the characteristic range of nonlocal effects, mediated by the molecular chain structure and length. However, in complex materials, such as polymers, the presence of random disorder further complicates such relation. Therefore, in the following analysis, we consider both cases where $range = L$ and where $range \neq L$.

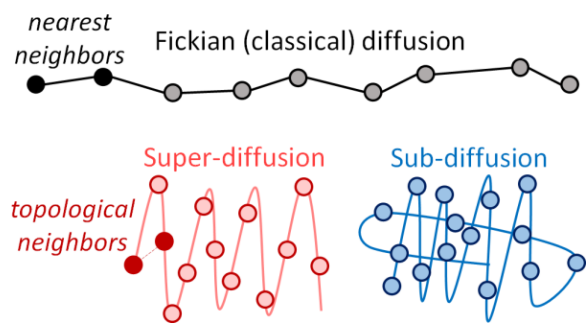

Fig. 3. Relation between molecular chain fold and the corresponding diffusion regime.

## IV. NUMERICAL RESULTS

In this section, we conduct a numerical experiment where the distance parameter in equation (6) is computed for various choices of disorder concentration $c$, three types of nonlocal interactions (determined by $s$), and scale $L$ of a reference lattice vector $\hat{v}$. Further details on the numerical simulation and criteria for interpreting numerical results can be found in Appendix A. The effect of numerical instabilities due to roundoff errors and other approximations is discussed in Appendix B. The loss of orthogonality during the Gram–Schmidt procedure is not considered likely to affect the present model, as discussed in detail in [15].

In this section, each plot $D_{s,\epsilon}^{\tau}(\tau)$ is an average of at least 10 realizations for the selected parameters, which minimizes inaccuracies due to the random realization of disorder values. Here, we are interested in the rate of decrease of each distance plot relative to the transport time within the medium. The total transport time is defined by the total number of timesteps, which is fixed to $\tau = 10,000$ in all simulations. In the following discussion, the phrase *rapidly drops to zero* refers to $D_{s,\epsilon}^{\tau}$ plots which drop to zero within the first 10-20% of the timesteps considered, and the phrase *slowly decreasing* refers to plots which do not drop to zero within the total number of timesteps considered.

The semi-crystalline polymer structure is modeled by the Hamiltonian in equation (1), with a discrete fractional Laplacian given in equations (2) and (3). Throughout this study, the concentration of disorder $c$ is allowed to vary in magnitude, but in all numerical experiments, the random disorder values $\epsilon_i \in [-c/2, c/2]$ are selected according to a uniform (flat) distribution. (See Kostadinova, et al. [63] for a study of transport in a 2D lattice, where disorder is selected from a Gaussian and a modified Gaussian distribution.) To model the different diffusion regimes resulting from nonlocal interactions, one can vary the fraction on the Laplacian in the interval $s \in (0,1)$ for super-diffusion, $s \in (1,2)$ for sub-diffusion, or fix it to $s = 1$ for normal (Fickian) diffusion. The present study is focused on three choices $s = 0.9, 1, 1.1$, which are considered representative for each regime. Once the distribution of disorder and fraction on the Laplacian are fixed, we vary the size $L$ of the reference lattice vector $\hat{v}$, which is given in equation (8).

To examine the dependence on disorder concentration $c$ in the three diffusion regimes, we initially fix the range of nonlocal interactions to $range = 300$ and consider only a reference vector of the same size $L = 300$. Figure 4 shows the time evolution of distance parameters, where the disorder concentration $c$ is varied across four scales from $c \sim 10^{-4}$ to $c \sim 10^{-1}$. In each plot, white dashed lines are used to distinguish among the four disorder regions. For the smallest examined disorder interval, $c \sim 10^{-4}$, the value of $D_{s,\epsilon}^{\tau}$ does not decrease appreciably from 1 in the super-diffusive regime (Fig. 4a), while substantial decrease is observed for the sub-diffusive case (Fig. 4c). The classical diffusion case shows a small decrease in $D_{s,\epsilon}^{\tau}$ for $c \sim 10^{-4}$, which is clearly enhanced in the range where $c \sim 10^{-3}$ (Fig. 4b). For all three diffusion regimes, the distance values drop rapidly to zero for the higher disorder ranges of $c \sim 10^{-2}$ and $c \sim 10^{-1}$, while preserving the expected overall behavior (i.e., sub-diffusive case decreases slower than diffusive, which decreases slower than super-diffusive).

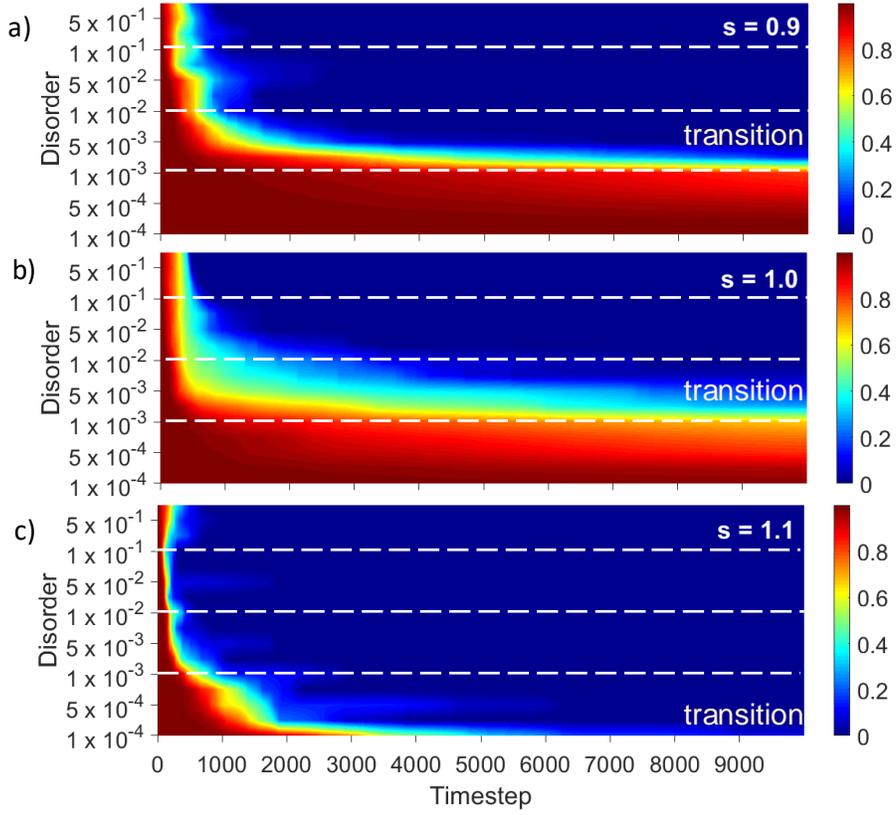

Fig. 4. Time evolution of distance parameter for various choices of disorder concentration in the a) super-diffusive, b) diffusive, and c) sub-diffusive regime. The color bar indicates the value of the computed distance parameter. In each plot, the dashed lines identify the four examined disorder scales.

The regions where the distance plots transition from slow decrease (de-localized behavior) to rapid drops are clearly visible in Fig. 4. The range $c \sim 10^{-3}$ seems transitional for both super-diffusive and diffusive realizations. In the sub-diffusive case, while all distance plots decrease to zero within the number of timesteps considered, the drops are smallest for disorder in the range $c \sim 10^{-4}$ (larger red region on the bottom left corner of Fig4c). Figure 5 shows time evolution of distance plots for representative values from these two smallest disorder scales: $c \sim 10^{-3}$ and $c \sim 10^{-4}$. In the smaller range of disorder, where $c \sim 10^{-4}$, for all examined cases the super-diffusive realizations decrease more slowly with time than the diffusive, which in turn decrease more slowly than the sub-diffusive ones. This is what we call the expected behavior (Fig. 5a). For values $c > 1 \times 10^{-3}$, we observe deviations from the expected behavior, in which both super- and sub-diffusive realizations decrease faster than the classical diffusion case (Fig. 5b). Since transport in a one-dimensional lattice is highly sensitive to concentration of disorder, higher concentrations lead to localization of all transport, even in the presence of long-distance interactions. In this case, the possibility for nonlocal interaction in the anomalous diffusion realizations does not produce qualitatively different transport behavior; instead, it allows the system to reach localization more rapidly. Since we are interested in regimes where qualitatively different diffusive transport is observable, the following analysis is focused on cases where $c \leq 1 \times 10^{-3}$.

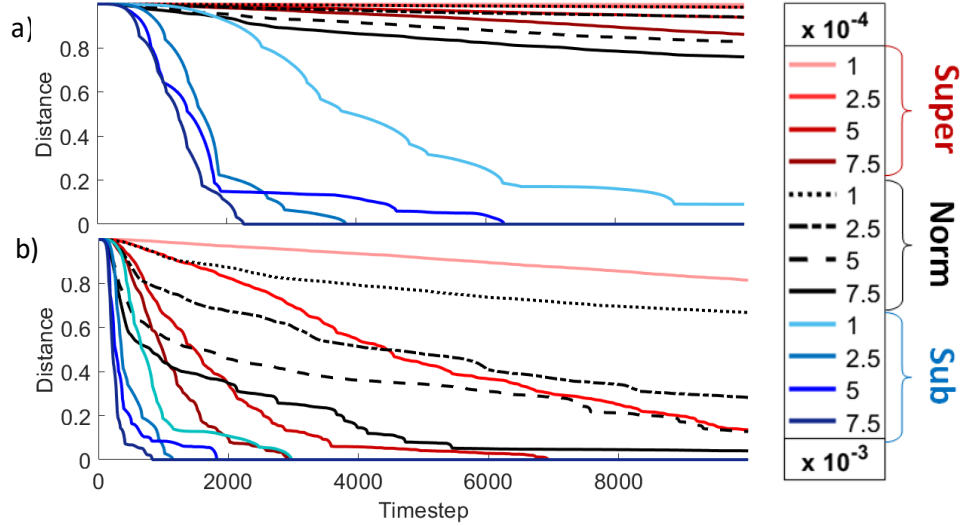

Fig. 5. Time evolution of distance parameter for super diffusion (red shades), normal diffusion (black), and sub-diffusion (blue shades) in the disorder range where a) $c \sim 10^{-4}$ and b) $c \sim 10^{-3}$.

To determine the expected size of crystalline regions in the semi-crystalline polymer, we fixed the concentration of disorder to $c = 1 \times 10^{-3}$ and examined how the distance plots change as a function of a reference lattice scale $L$ in the three diffusion regimes. As discussed in Sec. III, the physical interpretation of positive limiting value of $D_{S,\epsilon}^{\tau}$ as $\tau \to \infty$ is the existence of extended states, or transport, beyond a region of size comparable to the selected reference lattice vector size $L$. In other words, for a given choice of a Hamiltonian $H_{S,\epsilon}$, if $\lim_{\tau \to \infty} D_{S,\epsilon}^{\tau} > 0$, it is likely that transport is enhanced beyond the examined scale $L$, with the likelihood proportional to the limiting value of $D_{S,\epsilon}^{\tau}$. Conversely, if the distance rapidly drops to zero for a given $L$, it is expected that localization under the action of $H_{S,\epsilon}$ is enhanced at that scale. Thus, in the following analysis, we associate slowly decreasing distance plots with possible crystalline regions and rapidly decreasing plots with unlikely scales for the development of crystalline structures.

Assuming that the range of nonlocal interactions in the sub- and super-diffusive cases is comparable to the size of crystalline regions, in Fig. 6 we show the results for the three diffusion regimes at fixed disorder $c = 1 \times 10^{-3}$ and increasing size $L$ of the reference vector, while keeping $L = range$. The reference vector size was increased from $L = 50$ to $L = 1000$ in increments of $\Delta L = 50$. In the super-diffusive regime, the limiting $D_{S,\epsilon}^{\tau}$ values are positive for scales in the range $L \lesssim 450$ with a sharp transition region between nonvanishing and vanishing realizations (marked by the white dashed contour in Fig. 6a). The transition region for the classical Fickian diffusion case (white dashed contour in Fig. 6b) is much wider and smoother, which is to be expected from the local (nearest neighbor) interactions assumed in this calculation. In other words, for a fixed number of timesteps, the distance parameter approaches its limiting behavior more slowly in the classical diffusion calculation as compared to the anomalous diffusion cases, where long-distance interactions are assumed. Nevertheless, examination of Fig. 6 clearly shows that the assumed simulation size of $\tau = 10,000$ timesteps is large enough to reveal the expected differences in the transport behavior for the three regimes. Specifically, the transition region in the classical diffusion case is much broader and shifted to include smaller scales in the range $150 \leq L \leq 400$, where the limiting distance values decrease more rapidly as compared to the same scales in the super-diffusive case. Fig. 6c indicates that, for the selected conditions, transport is suppressed at all scales, which is expected for sub-diffusive behavior.

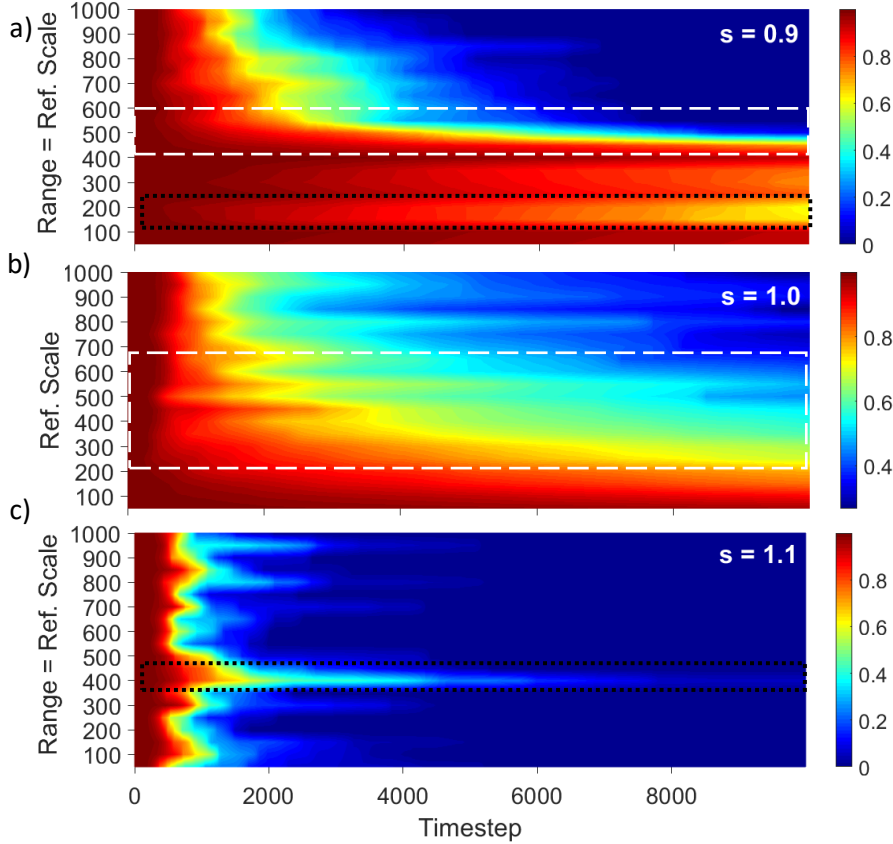

Fig. 6. Time evolution of distance parameter as a function of reference lattice scale in the a) super-diffusive, b) diffusive, and c) sub-diffusive regime. All calculations are performed for disorder $c = 1 \times 10^{-3}$ and assuming the range of nonlocal interactions is equal to the reference vector size. Regions where the distance plots transition from positive to vanishing limiting values are marked by a white dashed contour. Black dotted contours mark regions where the transport seems to deviate from the expected overall behavior.

Figure 6 also shows that both anomalous diffusion regimes exhibit realizations where transport seems to deviate from the overall behavior observed at the surrounding scales. Those realizations are marked by dotted black contours in Fig. 6 and correspond to $range = L = [150, 200]$ in the super-diffusive case and $range = L \sim 400$ in the sub-diffusive case. For each set of conditions, the time evolution of the distance parameter plotted in Fig. 6 is an average of 10 realizations, which minimizes fluctuations due to the random distribution of disorder. Thus, the observed unusual realizations may represent physically observable effects, resulting from the nonlocal character of the interactions. Possible numerical instabilities due to randomness of the disorder and numerical round-off errors are further discussed in Appendix B.

To explore the interplay between range of nonlocality and reference scale, we performed numerical experiments where $range$ is varied, while keeping the reference scale fixed at $L = 300$ and the disorder fixed at $c = 1 \times 10^{-3}$. The question of interest in such experiments is how the cutoff of nonlocality affects transport at a fixed scale of interest. Figure 7 shows the results for both anomalous diffusion cases, where the nonlocality influence was increased from $range = 50$ to $range = 1000$ in increments of $\Delta range = 50$. In both regimes, the realizations corresponding to unusual transport behavior (black dotted contours in Fig. 6a,c) are again present and even enhanced, as shown by the black dotted contours in Fig. 7. In addition, in the super-diffusive case (Fig. 7a), transport is impeded for $range = 50$, while in the super-diffusive case (Fig. 7b), transport is enhanced for $range = 50, 100, \& 150$. For both anomalous diffusion regimes, characteristic features of the results for larger range values are mostly unaffected as compared to the plots from Fig. 6. Additionally, the transition region in the super-diffusive case (white dashed contour in Fig. 7a) is preserved. This implies that, for fixed disorder concentration and fixed scale of interest $L$, the transport behavior can vary significantly for a range of nonlocal interactions $range \lesssim L$ that are smaller or comparable to the examined scale. However, as the nonlocality range exceeds the spatial scale $range > L$, transport is determined solely by the type of nonlocal interactions (super- or sub-diffusive).

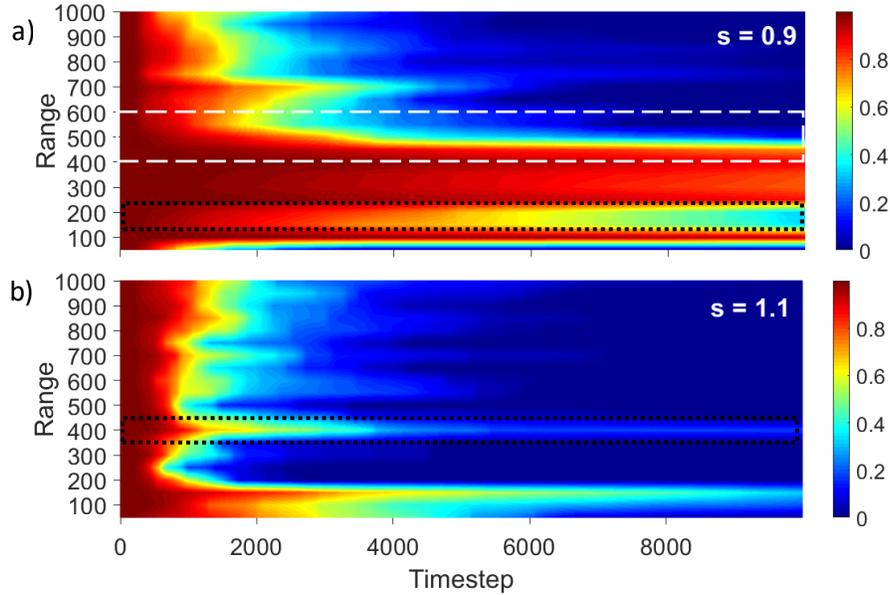

Fig. 7. Time evolution of the distance parameter for increasing range in the a) super-diffusive and b) sub-diffusive regimes. All calculations are performed for disorder $c = 1 \times 10^{-3}$ and assuming a fixed reference vector size $L = 300$. Transition regions are marked by a white dashed contour. Black dotted contours mark regions where the transport seems to deviate from the expected overall behavior.

To further examine these conclusions, we performed numerical experiments where the reference vector scale $L$ is varied, while keeping the range of nonlocal interactions fixed at $range = 300$ and the disorder fixed at $c = 1 \times 10^{-3}$. Here, the question of interest is how a given (fixed) range of nonlocal interactions affects transport at different spatial scales, including scales exceeding the range. Figure 8 shows the results for both anomalous diffusion cases, where the reference scale was increased from $L = 50$ to $L = 1000$ in increments of $\Delta L = 50$. For scales $L < 500$, which are comparable or smaller than the fixed influence $range = 300$, transport is enhanced in the super-diffusive case (Fig. 8a) and suppressed in the sub-diffusive case (Fig. 8b). As the reference scale increases, the transport behavior exhibits fluctuations, especially visible in the range $500 < L <$ 800 for the sub-diffusive realizations. This result is highly unexpected since it implies that extended states can exist in a disordered system, where negative nonlocal correlations induce anomalous diffusion.

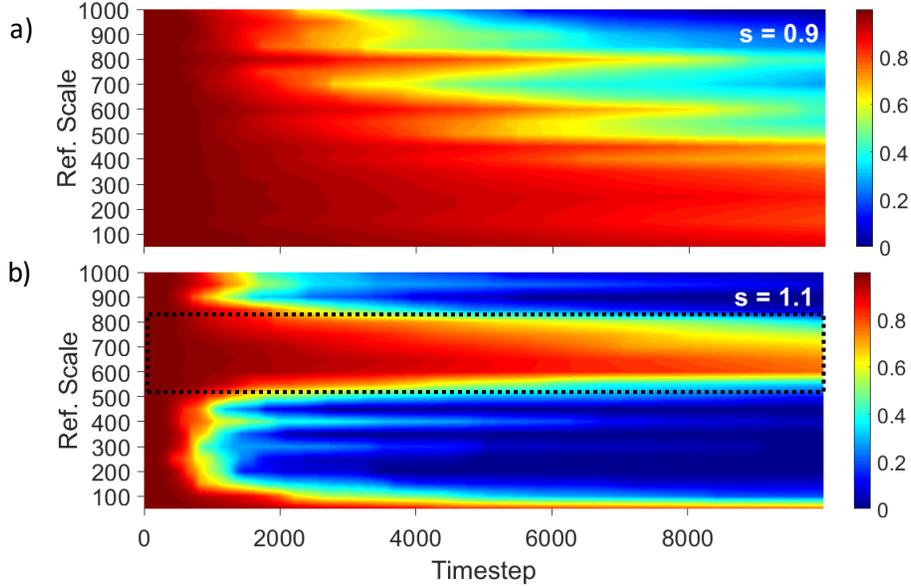

Fig. 8. Time evolution of the distance parameter for increasing reference vector scales in the a) super-diffusive and b) sub-diffusive regimes. All calculations are performed for disorder $c = 1 \times 10^{-3}$ and assuming a fixed range of nonlocal interactions $range = 300$. Black dotted contours mark the region, where the transport seems to deviate from the expected overall behavior.

## V. DISCUSSION

A possible physical explanation for the enhanced transport observed in Fig. 8b is the interplay between the two distinct localization mechanisms acting in this case: random disorder and sub-diffusive transport. In our model, sub-diffusion is represented by the fractional Laplacian $(-\Delta)^s$, with exponents $s \in (1,2)$, which results in non-linear transport (in the form of mean squared displacement that increases slower than linear with time). Numerical and experimental studies of one-dimensional granular crystals have recently demonstrated that disorder and non-linearity, which individually favor localization, can cancel out each other, yielding super-diffusive transport in the system [79], [80]. Based on an examination of the various possible sources of numerical errors affecting the computations (discussed in Appendix B), we conclude that the results observed in Fig. 8b are likely to have physical significance.

Initial insight into the mechanisms guiding transport in the sub-diffusion regime can be obtained by examination of Fig. 9, which provides a visual representation of the characteristic particle trajectories resulting from different choices of probability distribution functions corresponding to the three diffusion regimes. In each plot, the particle trajectories were *artificially generated* using numerical techniques from Tarantino et al. [81]. For Fickian (classical) diffusion (Fig. 9a), the probability distribution function (PDF) of particle positions approaches a Gaussian (Fig. 9e,f, black line). In the presence of memory (or correlations), the successive displacements of a walker are not fully independent, resulting in anomalous diffusion process with non-Gaussian PDF. Super-diffusion is commonly described as a random walk where the particles can make big jumps in displacement, called Lévy flights (Fig. 9b). Therefore, such a process can be characterized by a Lévy PDF (Fig. 9e,f, red line). In contrast, sub-diffusion is a process in which the walker's motion is impeded, leading to enhanced localization of the particle trajectories (Fig. 9c).

Since the driving mechanism, or generating process, for a sub-diffusive transport may be complex and highly dependent on the physical system, the corresponding PDF is not unique. However, it has been argued that irrespective of the generating law for a Markovian walk on an undirected network, the time-evolved probability distributions may converge only to two types of limiting distributions: namely, either Gaussian or Lévy distributions [82]. In other words, the asymptotic behavior is unaffected by any further complication or sophistication in the generating laws. To reflect this logic, we consider a case where the sub-diffusive trajectories were obtained using a PDF that is a superposition of truncated Gaussian and Lévy distributions. As can be seen in Fig. 9e,f, the resulting PDF is more concentrated at the origin, as compared to the Lévy distribution, yet it still exhibits "fat" tails further from the origin. Beyond a critical point, where the tails of the mixed distribution cross above the tails of the Gaussian ($\approx 4.3$ in Fig. 9f), nonlocal effects become important. To simulate highly localized trajectories, the tails of the mixed distribution can be truncated, such that the probability for long-distance jumps is minimized. In Fig. 9c, particle trajectories were generated from the mixed PDF truncated at $\pm 4$, which leads to suppression of the nonlocal effects and particle transport. Such a distribution is appropriate if one wants to model sub-diffusion as a process generated by increased waiting times between successive displacements.

In Fig. 9d, particle trajectories were generated from the same mixed distribution, but the truncation is performed beyond the critical point, at $\pm 100$. The resulting trajectories exhibit small-scale jumps which resemble the dynamics of the large-scale flights in Fig. 9b. However, the ensemble as a whole is localized in a small spatial region due to the presence of backward jumps, which can result in particles returning close to their original position. Although the data in Fig. 9 was artificially generated for representation purposes, the choice of a mixed (Gaussian + Lévy) PDF reflects the physical interpretation of the sub-diffusive fractional Laplacian used in the present study. As discussed in Sec. 2, Theorem 2 of [15], in the case $s \in (1,2)$, one can write $s = 1 + \tilde{s}$, where $\tilde{s} \in (0,1)$ and the sub-diffusive operator can be decomposed into $(-\Delta)^s = (-\Delta)^{\tilde{s}}(-\Delta)$. In other words, mathematically, sub-diffusion can be considered as a superposition of two successive actions (classical diffusion and super-diffusion). When these two actions are combined together, there is an enhanced probability that the particles jump backward toward the origin, leading to temporary localization in an area of space proportional to the scale of the jumps. In the absence of disorder, this temporary localization amounts to transport retardation when compared to the classical diffusion case.

In the 1D classical diffusion problem, the addition of random disorder introduces fluctuations of the corresponding probability distribution functions of particle displacement. Specifically, the probability for transport decreases exponentially with localization length, which is determined from the disorder concentration. In the super-diffusion case, the large-scale Lévy flights always act to enhance transport beyond the disorder-mediated localization length. In the sub-diffusion regime, one can expect that localization is enhanced when the scale of the jumps is of smaller or comparable size to the localization length. However, de-localization can also be achieved if the scale of the jumps exceeds the expected localization scales for a given disorder.

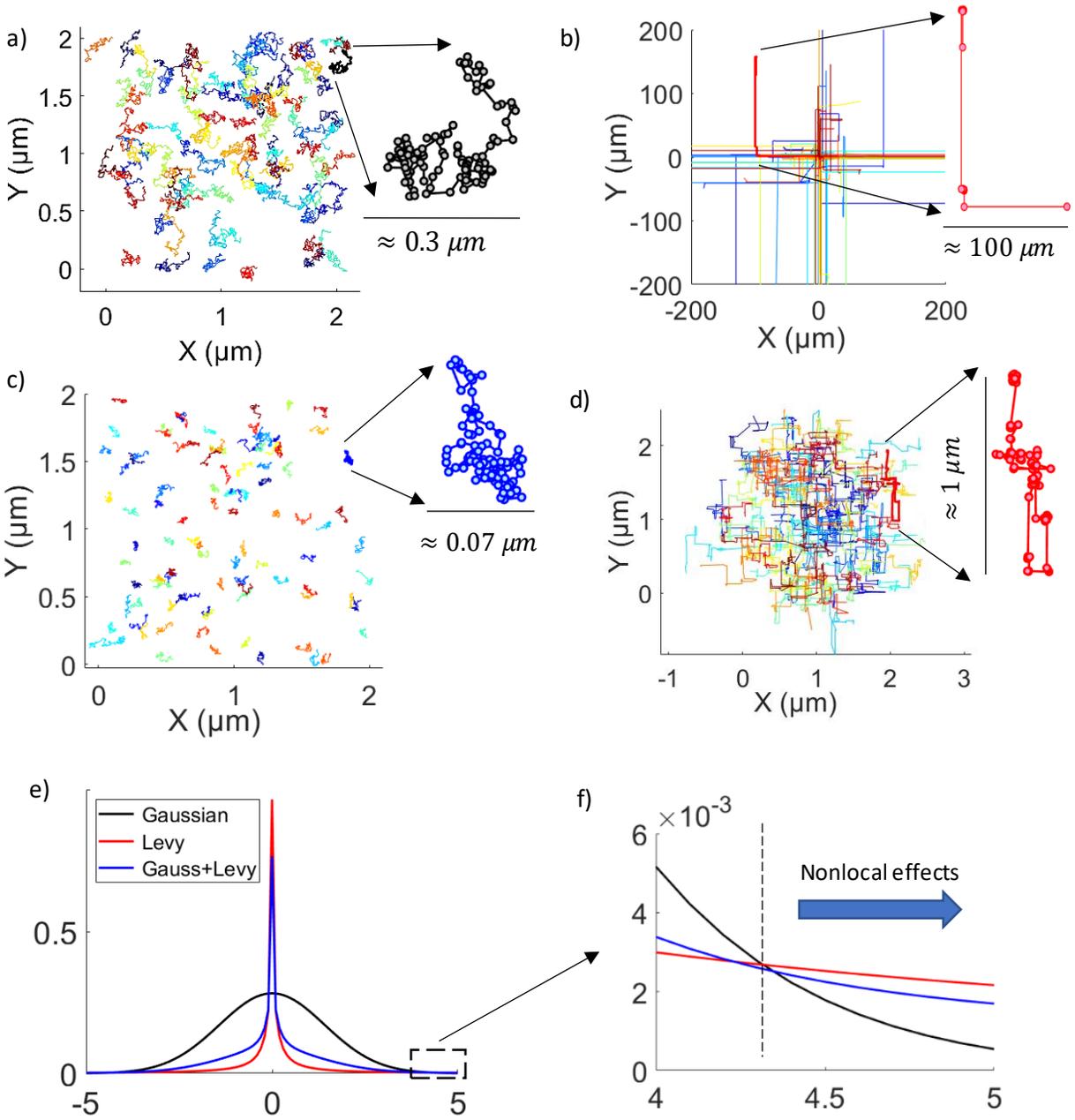

Fig. 9. Typical particle trajectories, representing a) Fickian diffusion, b) super-diffusion, and c) sub-diffusion. A Gaussian PDF was used to generate the trajectories in a), while a Lévy PDF was used in b). The trajectories in c) and d) were obtained from a superposition of Gaussian and Lévy PDFs with truncation at $\pm 4$ for part c) and at $\pm 100$ for part d). The three PDFs used to generate particle trajectories are superimposed in e). The tails of the three distributions are shown in f).

## VI. CONCLUSIONS AND FUTURE WORK

Herein we presented a numerical study of transport in disordered media, where nonlocal interactions can arise due to positive or negative correlations. The numerical experiments were performed using a Fractional Laplacian Spectral (FLS) model, which combines the spectral approach to the Anderson localization problem [47] with a series representation of the fractional Laplacian in 1D [15]. The resulting code models transport guided both by random disorder and by nonlocal interactions resulting in anomalous diffusion. As discussed in [47], [48], [64], the spectral approach provides knowledge of the energy states available to the system from iterative application of a Hamiltonian $H$, which does not require periodic boundary conditions or any up-scaling procedure. Applying boundary conditions to the domain of $H$ results in restriction of the Hilbert space and can lead to false information about the available energy states: specifically, the *a priori* exclusion of scattering states which belong to the absolutely continuous spectrum of the Hamiltonian. Using an up-scaling procedure, or averaging over subscale processes, can result in transport behavior of the system which exhibits *effective* long-range interactions. In contrast, the FLS technique employed in the present study models nonlocal interactions explicitly through the fractional Laplacian term and analysis of the resulting spectral structure of the Hamiltonian. This advantage of the technique makes it especially suitable for physical systems that are known to exhibit network-mediated interactions, such as polymers.

We developed a physical interpretation of the FLS model adapted for the study of light diffusion through a semi-crystalline polymer structure. A major finding of the numerical experiments is an unexpected transport enhancement at key spatial scales resulting from the interplay between competing localization effects. The existence of this effect has been recently discovered in experiments with 1D granular chains [79]. As polymers consist of large molecular chains, the presence of such anomalous transport may be related to key properties of the polymer structure, such as the formation of crystalline or amorphous regions resulting from particular folding sequences of the molecular chains. Thus, we expect that the findings of the present study can be extended and adapted to specific problems of interest within the greater scientific community, including research in electrorheological fluids, organic semi-conductors, proteins, and DNA.

As this is the first study applying the novel Fractional Laplacian Spectral technique to the problem of light transport in a polymer medium, our goal was to demonstrate the connection between the analytical formulation model, the numerical results, and their physical interpretation. Therefore, the focus of the present study is examination of the three qualitatively different transport regimes modeled by three choices for the fraction $s = 0.9, 1, 1.1$, which are considered representative for a super-diffusive, diffusive, and sub-diffusive regime, respectively. For each of the three $s$-values, we have varied disorder concentration (Fig. 4 and Fig. 5), range of nonlocal interaction (Fig. 6 and Fig. 7) and reference scale (Fig. 8), as well as further assessed the effects of randomness in the disorder (Fig. B1, and Fig. B2).

Preliminary results shown in Fig. 10 confirm that the qualitative diffusion properties established using the fixed choice $s = 0.9, 1, 1.1$ are preserved when the fraction is varied within the range $s \in (0,2)$ for the two scales of disorder considered $c = 10^{-4}$ and $c = 10^{-3}$. Namely, realizations where $s < 1$ consistently enhance transport, while cases where $s > 1$ enhance localization when compared to the classical case $s = 1$. However, we do not expect that the regimes where $s < 1$ and $s > 1$ have a symmetric effect on the transport properties. Therefore, while the case $s = 0.5$ is often recognized as representative of super-diffusive behavior, the case $s = 1.5$ should not be necessarily assumed as typical for the sub-diffusive regime. This becomes apparent from the discussion in Sec. V, where we suggested that in the case $s \in (1,2)$, one can write $s = 1 + \tilde{s}$, where $\tilde{s} \in (0,1)$ and the sub-diffusive operator can be decomposed into $(-\Delta)^s = (-\Delta)^{\tilde{s}}(-\Delta)$. As the value $\tilde{s}$ is increased, the probability for nonlocal jumps is increased, which can diminish the localization effect expected in that regime. As this point is subtle, it will be further discussed in our future work.

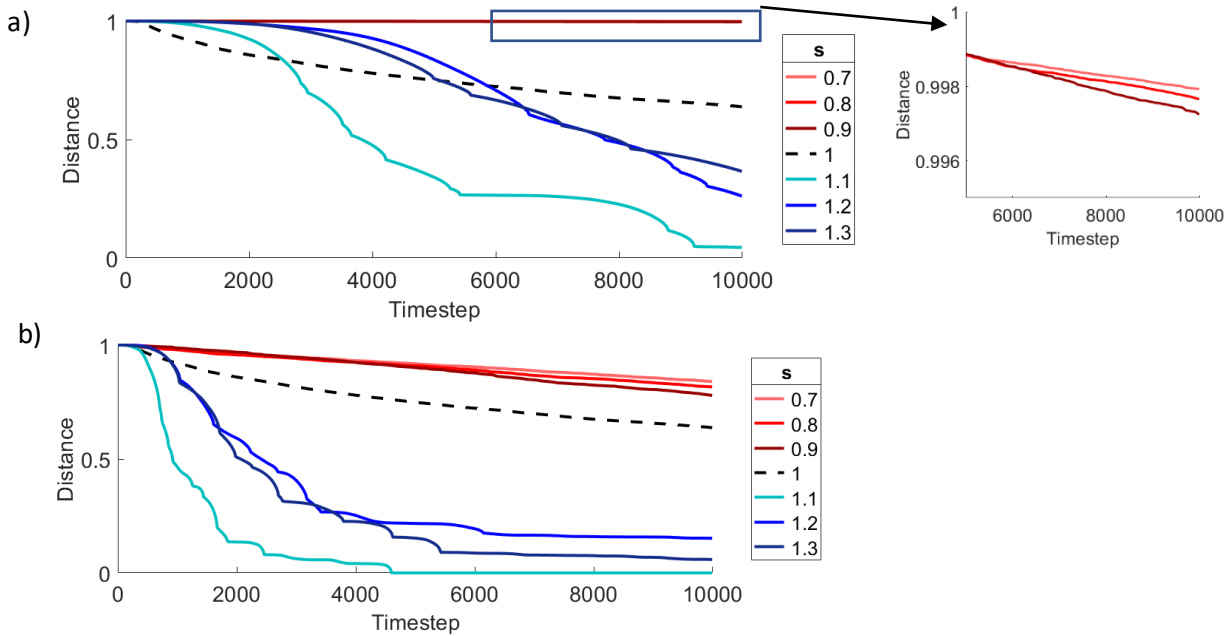

Fig. 10. Time evolution of distance parameter for various fractions in the range $s \in (0,2)$ for disorder a) $c = 10^{-4}$ and b) $c = 10^{-3}$. The three diffusion regimes are colored as follows: super diffusion (red shades), normal diffusion (black), and sub-diffusion (blue shades).

Another direction of future work includes testing the predictions of the FLS models in numerical and laboratory experiments with multi-chain dusty plasmas. These are arrangements of charged grains into string-like structures, which can be thought of as macroscopic analogues to molecular chains in polymers. Dust grains immersed in plasma become negatively charged and are subject to both ion drag forces and collective (nonlocal) interactions. As a result, dusty plasmas can self-organize into strongly coupled fluids with electrorheological properties, i.e., the ability to undergo homogeneous-to-string structural transitions when acted upon with electric fields [83], [84]. More importantly, the dust particles are directly observable with a video camera, which makes them ideal for the study of self-organization and stability, phase transitions, and transport phenomena at the kinetic (individual particle) level. This allows for the direct examination of anomalous diffusion as a function of scale and range of nonlocal interactions. The former is defined by the number of particles forming chains in the dusty plasma liquid, while the latter can be varied by tuning the plasma parameters in the experiment.

ACKNOWLEDGMENTS

This work was supported by the NSF grant numbers 1903450 (EGK, JLP, CDL, and LSM), 1707215 (LSM and TWH), and 1740203 (TWH and LSM), NSF-DMS grant number 1802682 (CDL), and NASA grant number 1571701 (TWH and LSM).

Since August 2020, CDL has been serving as a Program Director in the Division of Mathematical Sciences at the National Science Foundation (NSF), USA, and as a component of this position, she received support from NSF for research, which included work on this paper. Any opinions, findings, and conclusions or recommendations expressed in this material are those of the authors and do not necessarily reflect the views of the National Science Foundation.

## Appendix A. DETAILS OF THE NUMERICAL SIMULATION

Consider the *random fractional discrete Schrödinger operator* in equation (1), which we repeat here for convenience:

$$H_{s,\epsilon} := (-\Delta)^s + \sum_{i \in \mathbb{Z}} \epsilon_i \langle \cdot, \delta_i \rangle \delta_i , \tag{A1}$$

with a fractional Laplacian $(-\Delta)^s$ given by the series representation in equations (3) and (4) and random variables $\epsilon_i \in [-c/2, c/2\,]$, where the concentration of disorder $c$ is a nonzero positive number. The $\{\delta_i\}$ are the standard basis vectors of the 1D integer space $\mathbb{Z}$, i.e. each $\delta_i$ corresponds to a discrete point in the 1D lattice space. The size of the 1D simulation $N_{tot}$ is determined by the number of timesteps $N_\tau$ and number of neighbors contributing to the nonlocal interactions at each timestep, which we call *range* in the discussion. In other words, the lattice size is given by

$$N_{tot} = 2(range)N_\tau + 1. \tag{A2}$$

In the classical case where $s = 1$, the assumed interactions are local and $range = 1$ since interactions with only one nearest neighbor in each direction are considered. In each numerical experiment, we initially fix $c$ and then obtain a computer-generated realization of the random variables $\epsilon_i$ according to a flat distribution in the interval $[-c/2, c/2]$. As the variables $\epsilon_i$ are independent and identically distributed in this interval, they represent uncorrelated disorder, which affects the interactions only locally. As given realization of $\epsilon_i$ is kept fixed throughout the corresponding numerical experiment, the disorder-induced effects are also time-independent.

The value of 1 is then assigned at the midpoint of the lattice, which corresponds to letting the initial state $\varphi_0 \equiv \delta_0$. Next, we calculate the values of $D_{s,\epsilon}^\tau$ for $n \in \{0,1,2,...,\tau\}, \tau \in \mathbb{N}$, for each fraction $s$ using equation (6). Throughout this paper, we focused on three choices $s = 0.9, 1, 1.1$, which are considered representative for a super-diffusive, diffusive, and sub-diffusive regime, respectively. Finally, the calculated distance values $D_{s,\epsilon}^n$ are plotted to determine their limiting behavior as $\tau \to \infty$.

According to the spectral approach outlined in Sec. II, the spectrum of the Hamiltonian in equation (1)/(A1) includes an absolutely continuous (ac) part if one can demonstrate that $\lim_{\tau \to \infty} D_{s,\epsilon}^\tau > 0$. The presence of ac spectrum can then be interpreted as existence of extended (scattering) states. Due to the finite nature of the simulation and the probabilistic nature of random disorder, rescaling and function fitting can be employed to determine the limiting value of the distance parameter. Since $D_{s,\epsilon}^n$ is a positive, monotonically decreasing sequence, one can construct approximate lower bounds (with respect to the probability distribution) of $\lim_{\tau \to \infty} D_{s,\epsilon}^\tau$. In realizations of the experiment where $\lim_{\tau \to \infty} D_{s,\epsilon}^\tau$ does not drop rapidly to zero (e.g., Fig. 5a, super-diffusion realizations), the distance parameter can be plotted on a log-log scale to distinguish rapidly-decreasing from slowly-decreasing realizations and the corresponding transport behavior. As described in [15], Sec. 4.2, the following criterion can be used for interpretation of the qualitative differences in the plots:

**Numerical Interpretation Criterion 1:** For a fixed realization of $\epsilon_i$ and reference vector $\hat{v}$, and for sufficiently large number of timesteps $\tau$, if $D_{s2,\epsilon}^\tau > D_{s1,\epsilon}^\tau$, and $H_{s1,\epsilon}$ is known to exhibit extended states, then the operator $H_{s2,\epsilon}$ also exhibits extended states. Similarly, if $D_{s2,\epsilon}^\tau < D_{s1,\epsilon}^\tau$, and $H_{s1,\epsilon}$ is known to exhibit localized states, then the operator $H_{s2,\epsilon}$ also exhibits localized states.

Since it is known that for $s = 1$ in 1D all energy states resulting from the operator in (1)/(A1) are localized [85], the distance $D_{1,\epsilon}^\tau$ limits to zero for all nonzero realization of disorder and all choices of a reference vector $\hat{v}$ (as discussed in Sec. 6.2 of [47]). Thus, this case can be used as a reference when interpreting the more complex cases resulting from $s \neq 1$ in the interval $s \in (0,2)$:

**Numerical Interpretation Criterion 2:** For a fixed realization of $\epsilon_i$ and reference vector $\hat{v}$, and for sufficiently large number of timesteps $\tau$, if $D_{s\neq 1,\epsilon}^\tau > D_{1,\epsilon}^\tau$, then the operator $H_{s\neq 1,\epsilon}$ is less likely to exhibit localization than the operator $H_{1,\epsilon}$. Similarly, if $D_{s\neq 1,\epsilon}^\tau < D_{1,\epsilon}^\tau$, then the operator $H_{s\neq 1,\epsilon}$ is more likely to exhibit localization than the operator $H_{1,\epsilon}$

Note that this second criterion is only valid for small concentrations of disorder ($c = 10^{-4} - 10^{-3}$ in this paper) where we expect that the scattering effect of nonlocality is comparable to the localizing effect of disorder, which suggests that it is possible for extended states to occur. For higher values of $c$ the disorder-induced localization behavior is expected to dominate the dynamics. In those cases, the nonlocal nature of the interactions induced by $H_{s\neq 1,\epsilon}$ can actually cause the plot of $D_{s\neq 1,\epsilon}^\tau$ to decrease more rapidly to zero than the plot of $D_{1,\epsilon}^\tau$, where the dynamics are guided by the nearest-neighbor interactions. In other words, for this strongly localized regime, at each time step the calculation of $D_{s\neq 1,\epsilon}^\tau$ includes contributions of many more neighbors and their corresponding on-site disorder than the calculation of $D_{1,\epsilon}^{\tau}$, which includes contributions from only the two nearest neighbors at a time.

## Appendix B. VALIDATION OF NUMERICAL RESULTS

Section IV presented numerical experiments studying transport in disordered media where nonlocal interactions can yield anomalous diffusion. The results demonstrated the possibility of transport enhancement in the regime where different localization mechanisms compete and effectively cancel each other out. As these results may have significant physical applications, in this section we examine possible sources of numerical errors affecting the computation. Specifically, we focus on i) statistical fluctuations due to averaging over random realizations of disorder and ii) the effect of truncation on the validity of asymptotic expressions. Instabilities due to loss of orthogonality in the Gram Schmidt procedure employed in the calculations are expected to have negligible effect on the results, as discussed in Sec. 4.3 of [15].

### 1. Randomness of Disorder

Due to the 1D nature of the simulation, it is expected that even small concentrations of random disorder can greatly affect the observed transport behavior. In Sec. IV, we saw that for small disorder concentrations $c \sim 10^{-4}$, the distance plots follow the expected behavior: they decrease slower in the super-diffusive regime and faster in the sub-diffusive regime when compared to the classical diffusion case. However, once a certain disorder threshold is crossed (resulting in strong localization behavior) it was observed that both anomalous diffusion cases yield distance parameters that decrease more rapidly to zero than the classical case (Fig. 5). The deviations from the expected behavior for $c > 10^{-3}$ are due to the nonlocal nature of the fractional Laplacian, which allows for the propagation of "localization information" at a faster rate as compared to the classical case, where the same process is guided by nearest-neighbor interactions. Thus, at high concentrations of disorder, the anomalous diffusion distance plots decay more rapidly in response to the strong localization effect. We expect that this phenomenon would be less prominent in higher dimensions. However, for the 1D simulations of Sec. IV, we set a threshold of $c = 1 \times 10^{-3}$ as the maximum disorder concentration for which the diffusion dynamics is not entirely dominated by disorder-driven localization.

Even at $c = 1 \times 10^{-3}$, the role of random disorder may have significant effect on the results, which is especially important in light of the unexpected de-localization of distance plots observed for scales $500 < L < 800$ in the sub-diffusive regime (Fig. 8b). To minimize the role of random disorder, we performed numerical experiments with the same parameters as in Fig. 8 but using one order of magnitude smaller disorder $c = 1 \times 10^{-4}$. The results presented in Fig. B1 demonstrate that although transport is naturally enhanced for all realizations in the smaller disorder case, the region $500 < L < 800$ in the sub-diffusive regime (Fig. B1b) still exhibits the unexpected behavior observed in Fig. 8b. This suggests that the unusual transport behavior is not caused by a random effect due to the disorder distribution.

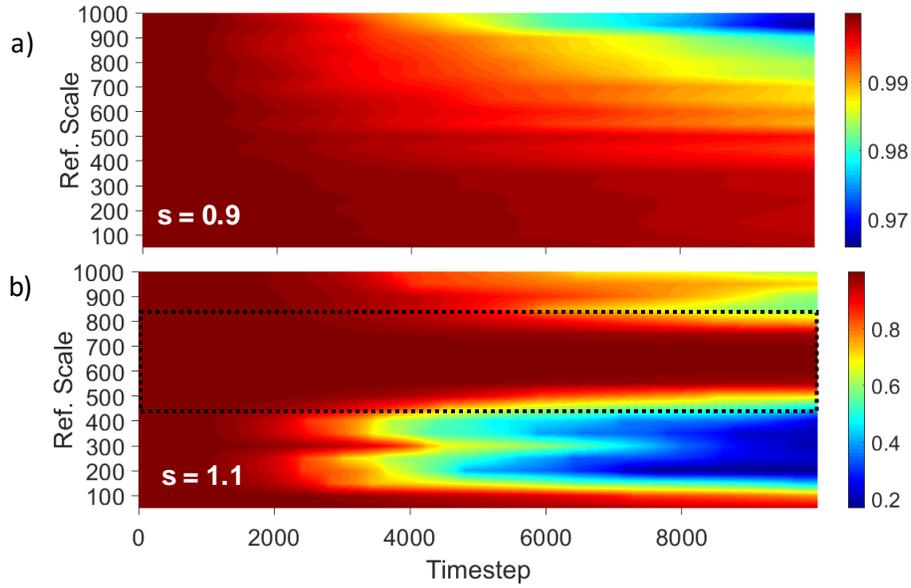

Fig. B1. Time evolution of the distance parameter for increasing reference vector scales in the a) super-diffusive and b) sub-diffusive regimes. All parameters are the same as in Fig. 8 but computed at smaller disorder $c = 1 \times 10^{-4}$. Black dotted contour marks the region where the transport deviates from the expected overall behavior.

Another test for the role of randomness is to examine how improved averaging affects the outcome of the simulation. In the previous section, we mentioned that each distance plot is an average of 10 realizations of the same numerical experiment, which is needed to minimize fluctuations due to the random realization of the disorder in each individual run. Figure B2 shows the time evolution of the distance plots for the same parameters as in Fig. 8 in the range $400 \leq L \leq 800$ computed using 10 realizations (Fig. B2a) and using 50 realizations (Fig. B2b). As the number of realizations is increased, the enhanced transport behavior is preserved and identified more clearly in the region $550 \leq L \leq 750$. This suggests that averaging over 10 realizations is sufficient to identify interesting features of the transport behavior, which ensures reasonable computation times. However, averaging over larger number of realizations should be used for a more detailed study of interesting regions.

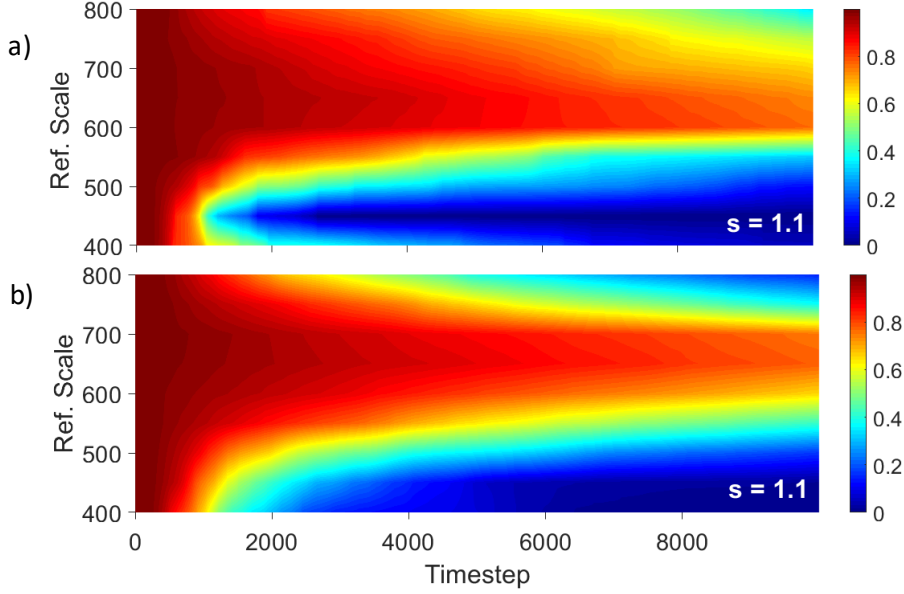

Fig. B2. Time evolution of the distance parameter in the sub-diffusive case for increasing reference scales in the range $400 \leq L \leq 800$ calculated by averaging a) 10 realizations and b) 50 realizations for each set of parameters. All calculations are performed for disorder $c = 1 \times 10^{-3}$ and assuming a fixed range of nonlocal interactions $infl = 300$.

## 2. Truncation Effects

In addition to randomness, the accuracy of the numerical results can be affected by truncation due to the selected range of nonlocal interactions. In Sec. II, we introduced the series representation of the fractional Laplacian in equation (2), which can be rewritten as

$$(-\Delta)^s u_n = \sum_{m \in \mathbb{N}} (2u_n - u_{n-m} - u_{n+m}) K_s(m), \tag{B1}$$

where the kernel $K_s$ is given by equation (3). This form of the equation is equivalent due to the symmetry of the kernel. Although in the analytical expression the summation goes to infinity, in the numerical simulation the explicit summation is performed over a finite number of terms, determined by the selected $range$ of nonlocal interactions. Specifically, the numerical equivalent of equation (B1) becomes

$$(-\Delta)^s u_n = \sum_{m=1}^{range} (2u_n - u_{n-m} - u_{n+m}) K_s(m) + R_{range}(u_n), \tag{B2}$$

where the remainder is given by

$$R_{range}(u_n) = \sum_{m=range+1}^{\infty} (2u_n - u_{n-m} - u_{n+m}) K_s(m) \sim \frac{1}{(range)^2}. \tag{B3}$$

As discussed in Sec. 4 of [15], the remainder is inversely proportional to the square of the $range$. In the present study, the remainder $R_{range}$ is discarded. Thus, we expect that the numerical results increase in accuracy with increasing $range$. The smallest influence value used in the presented numerical experiments is $range = 50$, which yields the largest remainder $R_{50} \sim 4 \times 10^{-4}$. All other results were obtained using larger $range$ leading to smaller remainder, with the most common choice being $range = 300$, which yields $R_{300} \sim 10^{-5}$. Even though the remainders are rather small, we acknowledge the possibilities of numerical instability artifacts due to the large number of timesteps used, $\tau = 10{,}000$.

Deviations from the expected transport behavior may also correspond to some analytical feature of the model. Since expressions related to the fractional Laplacian are exact in the asymptotic limit, it is expected that larger values for $range$ (smaller truncation) yield improved results, assuming other numerical instabilities are minimized. For smaller $range$ (larger truncation), the calculation using equation (B2) may not be accurate, which would explain the small fluctuations observed in Figs. 6 and 7 (marked by black dotted contours). However, the unusual de-localization in the sub-diffusive regime in Fig. 8b occurs at large values of $range$, which are expected to yield more accurate results. Finally, the sub-diffusive regime is not as well-behaved as the super-diffusive case due to the associated differential operator lacking a so-called maximum principle. Therefore, further validation of the present numerical results will be sought by exploring alternative implementation methods and comparing with experimental results.